# Active silicon integrated nanophotonics: ferroelectric BaTiO$_3$ devices

Chi Xiong[1], Wolfram H. P. Pernice[1,†], Joseph H. Ngai[2], James W. Reiner[2], Divine Kumah[2], Frederick J. Walker[2], Charles H. Ahn[2,3] and Hong X. Tang[1,*]

[1]Department of Electrical Engineering, Yale University, New Haven, CT 06511, USA

[2]Department of Applied Physics, Yale University, New Haven, CT 06511, USA

[3]Department of Mechanical Engineering and Materials Sciences, Yale University, New Haven, CT 06511, USA

† Current address: Institute of Nanotechnology, Karlsruhe Institute of Technology, 76344 Eggenstein-Leopoldshafen, Germany

*Correspondence author electronic mail: hong.tang@yale.edu

**Abstract:** The integration of complex oxides on silicon presents opportunities to extend and enhance silicon technology with novel electronic, magnetic, and photonic properties. Among these materials, barium titanate (BaTiO$_3$) is a particularly strong ferroelectric perovskite oxide with attractive dielectric and electro-optic properties. Here we demonstrate nanophotonic circuits incorporating ferroelectric BaTiO$_3$ thin films on the ubiquitous silicon-on-insulator (SOI) platform. We grow epitaxial, single-crystalline BaTiO$_3$ directly on SOI and engineer integrated waveguide structures that simultaneously confine light and an RF electric field in the BaTiO$_3$ layer. Using on-chip photonic interferometers, we extract a large effective Pockels coefficient of 213 ± 49 pm/V, a value more than six times larger than found in commercial optical modulators based on lithium niobate. The monolithically integrated BaTiO$_3$ optical modulators show modulation bandwidth in the gigahertz regime, which is promising for broadband applications.

The epitaxial growth of perovskite oxides on Si enables new functionalities such as piezoelectricity[1] and ferroelectricity[2] to be integrated with Si technology. Among the class of perovskites, barium titanate ($BaTiO_3$ or BTO) is a prototype functional oxide which has traditionally been used in ceramic capacitors and recently been considered for advanced applications in microelectronics[3, 4] and optics[5, 6]. For optical applications, barium titanate is attractive for several reasons. BTO thin films are transparent to light at optical frequencies[7], and the index of refraction is highly sensitive to applied electric fields[8, 9]. Moreover, BTO can be grown as an epitaxial single crystal on the Si (100) surface[10, 11], making it an ideal material for integrated optical modulators. While the epitaxial of BTO has been previously demonstrated, significant material science and device integration challenges have to be overcome to translate the material advantage into a real device benefit. A key breakthrough will be to develop integrate planar waveguide devices in BTO, which can be monolithically integrated with existing silicon photonics and microelectronic components[12, 13].

Here, we demonstrate silicon integrated planar lightwave devices incorporating ferroelectric BTO as an active waveguide material. We use reactive molecular beam epitaxy (MBE) to grow single crystal BTO films as thick as 80 nm on silicon-on-insulator (SOI) substrates using a thin buffer layer of strontium titanate (STO). The use of SOI substrates ensures full compatibility with existing silicon-based photonics and microelectronics technologies. Besides challenges in material development, one practical barrier for implementing BTO in photonic devices is its index of refraction, which is lower than that of Si. To guide light in the BTO layer, we design and fabricate a horizontal slot waveguide structure[14, 15], which sandwiches the BTO in between two high index silicon layers. This geometry increases the optical field in the BTO and results in a strong overlap of optical and electrical fields with significant electro-optic coupling. Using this geometry, we create lightwave circuits on BTO-on-silicon substrates, including nanophotonic components, such as microring resonators and optical interferometers. Using a Mach-Zehnder interferometric device, we extract a large effective electro-optic coefficient of 213 ± 49 pm/V in the BTO thin films. Measurements of the electro-optic effect demonstrate the poling of the ferroelectric domains in our high quality epitaxial films. With this approach, we realize electrically driven Mach-Zehnder interferometers and microracetrack resonators that enable electro-optic modulation up to 4.9 GHz, a value that is only limited by the device drive circuit. Our demonstration of BTO as an active optical element monolithically integrated on silicon opens new avenues toward the application of ferroelectric oxides in fully integrated silicon photonic circuits.

Compared to silicon, BTO has a lower refractive index ($n_{BTO}$ = 2.38, $n_{Si}$=3.47 @ 1550 nm). As a result, conventional rib and strip waveguides prevent optical confinement in the BTO layer. To circumvent this issue, we design a special horizontal slot waveguide consisting of a thin BTO layer sandwiched between two high-index silicon layers, as shown in Figure 1 (a). The undoped device layer of the SOI substrate serves as the lower Si layer, while an equally thick amorphous layer of undoped silicon serves as the upper Si layer. Plasma-deposited amorphous Si (a-Si) can support low-loss optical transmission at telecommunication wavelengths[16, 17]. The waveguide is lithographically patterned into the amorphous silicon layer, circumventing the challenge of dry etching BTO. As shown in Figure 1 (b), we use finite element analysis to determine the mode properties of the waveguide and static electric field distribution when a bias voltage is applied to the electrodes. The simulated structure consists of a 110 nm thick crystalline silicon layer, an 80 nm BTO layer, and a 110 nm a-Si layer. The waveguide width is optimized to be 800 nm to allow for single-mode operation. The simulation demonstrates that the fundamental TE-like optical mode is well confined in the a-Si-BTO-Si sandwich structures, as shown in Figure 1 (b). In particular, strong overlap between the optical field and electric field is found within the BTO layer. The effective mode index is numerically calculated to be $n_{\text{eff}} = 2.60$. Figure 1 (c) shows an optical micrograph of a fabricated BTO racetrack modulator. The simulated parameters are matched to the fabricated devices, as described below.

To exploit the electro-optic effect of the BTO waveguide, we apply the electric field through electrodes patterned in the vicinity of the waveguide. This arrangement allows the waveguide structure to be employed as a modulator by exploiting the electro-optic modulation of the effective refractive index inside the horizontal slot. This type of partially etched horizontal slot waveguide is ideally suited for the realization of planar electro-optic modulators because it leads to high confinement of both optical and electric fields in the sandwiched slot layer. The zoom-in scanning electron micrograph in Figure 1 (d) shows that the gold electrodes (in yellow) are separated from the a-Si waveguide (in blue) by 1 μm. It is worth noting that only a small portion of the optical mode is found outside the ridge structure. This strong confinement is important to prevent optical absorption in the gold electrodes, permitting a small gap between the gold electrodes and the waveguide and higher RF fields. To maximize the modulation efficiency, it is important to maximize the overlap between the guided light and RF field, which is quantified by the electro-optic integral $\Gamma = \dfrac{g}{V} \dfrac{\iint E_{x,op}^2 E_{Mod} \, dxdz}{\iint E_{x,op}^2 \, dxdz}$, where $g$ is the electrode gap, $V$ is the voltage on the electrodes, $E_{x,op}$ is the optical field, and $E_{Mod}$ is the RF field. Figure 1 (e) plots the simulated electro-optic integral as a function of the a-Si thicknesses for different BTO thickness. According to the simulations, the electro-optic integral $\Gamma$ increases significantly as the BTO thin film thickness increases. In this work, to ensure high yield of the epitaxial BTO thin films, a BTO thickness of 80 nm is chosen. The electrode gap is fixed at 1 μm, which ensures that the optical loss due to absorption in the electrodes is negligible.

At room temperature (T = 300 K), BaTiO3 is tetragonal with bulk lattice constants of $a = b = 3.992$ Å, $c = 4.036$ Å and a ferroelectric polarization aligned along the c-axis[18]. The $a/b$ lattice constant of bulk BTO differs from the (100) surface lattice constant of silicon (~3.84 Å) by nearly 4%. Thus, the crystallization and growth of high quality single crystal BTO directly on Si (100) is difficult. To mitigate this mismatch in lattice constants, an 8 nm (20 unit cell) thick buffer layer of SrTiO3 (STO) is first grown on the Si. Subsequently, an 80 nm thick layer of BTO is grown on the STO buffer.

Figure 2 (a) shows a typical X-ray diffraction (XRD) θ-2θ scanning measurement of the BTO films. From the measurement of eight in- and out-of-plane Bragg reflections, we determine that the BTO film is tetragonal with an out-of-plane lattice constant of 3.998 Å and an average in-plane lattice constant of 4.03 Å. The full width at half maximum of the BTO (200) rocking curve Δω is 0.5 degrees (shown in the right inset of Figure 2 (a)). The left inset in Figure 2 (b) shows an atomic force micrograph of the surface of the 80 nm-thick BTO layer. Except for a few islands, the BTO surface has a root-mean-square (*rms*) roughness as low as 0.4 nm. The footprint of the islands is about 50 nm (W)×10 nm (L)×30 nm (H), which is small compared with the optical wavelength of concern (1550 nm), and hence the light scattering caused by these scarce defects is virtually negligible. Figure 2 (b)-(d) shows several transmission electron micrographs of the cross section of the BTO horizontal slot waveguide. From the micrographs, the four essential elements of the waveguide structure can be seen, namely the crystalline and amorphous silicon layers, the BTO, and the SiO2 layer. The thin interfacial SiO2 layer arises from the post-growth anneal in oxygen at 800 °C, which was performed to eliminate residual oxygen vacancies.

From the XRD results, the BTO layer is relaxed relative to the SOI substrate. Thus, as the BTO is cooled from growth temperatures, the SOI substrate imparts tensile strain to the relaxed BTO film, since the former has a much smaller coefficient of thermal expansion than the latter. Consequently, the in-plane lattice parameter of the BTO is larger than the out-of-plane lattice parameter, leading to a ferroelectric polarization that lies in plane with a domain structure having four variants, with their polarizations oriented 90° to one another. This in-plane orientation of the spontaneous polarization is important as it allows the use of planar electrodes to induce electro-optic modulation, as described below.

For use in electro-optically active devices the field orientation provided by the driving electrodes needs to be properly chosen. The change in refractive index $n$, of single crystal BTO to an applied electric field, $E$, is described by the following reduced tensor:

$$\begin{bmatrix} \Delta(1/n^2)_1 \\ \Delta(1/n^2)_2 \\ \Delta(1/n^2)_3 \\ \Delta(1/n^2)_4 \\ \Delta(1/n^2)_5 \\ \Delta(1/n^2)_6 \end{bmatrix} = \begin{bmatrix} 0 & 0 & r_{13} \\ 0 & 0 & r_{13} \\ 0 & 0 & r_{33} \\ 0 & r_{42} & 0 \\ r_{51} & 0 & 0 \\ 0 & 0 & 0 \end{bmatrix} \cdot \begin{bmatrix} E_x \\ E_y \\ E_z \end{bmatrix},$$

where the $r_{mn}$ coefficients describe the electro-optic response for tetragonal BTO. Values for the electro-optic coefficients of BTO have been reported[19] to be as high as $r_{33} = 105 \pm 10$ pm/V and $r_{51}=r_{42}=1300\pm100$ pm/V. According to Fig. 1(e), the electro-optic integral $\Gamma$ for our adopted waveguide geometry is about 22%. As a result, 1400×22% = 308 pm/V will be an upper bound of the EO coefficient in our waveguide geometry. The largest electro-optic coefficient, $r_{51}$, of BTO structures is used when the direction of the applied electric field is oriented along a [100] direction perpendicular to the polar axis of ferroelectric BTO and the optical polarization of the guided light lies along a [011] direction, which is 45° to the ferroelectric polarization direction. For a channel waveguide, this arrangement of optical and electric fields is difficult to realize. However, an optimal electro-optic coupling can be achieved for a device that guides a TE-like mode along the [011] direction of the BTO. The relations between the ferroelectric polarization, waveguide and E-field orientation are illustrated in the schematic in the inset of Fig.1(c). In this configuration, the optical field is polarized along the [0$\bar{1}$1] direction, with a component of the electric field aligned along the [010] direction of BTO. In this configuration, all the four domain variants of the BTO (illustrated in inset of Fig.1(c)) can be poled by an applied DC electric field and contribute to the effective electro-optic coefficient via the large $r_{51}$ coefficient.

The first photonic component we develop for integrated BTO photonic circuits using the process flow shown in Fig. 3(a) is a focusing grating coupler[20], which is used to couple light in and out of the chip. The gratings are etched only in the a-Si layer and thus form a shallow grating, which offers improved coupling efficiency[21]. With optimized filling factor and grating period, a coupling loss of -7 dB at the central coupling wavelength is achieved.

We next fabricate on-chip waveguides in a Mach-Zehnder interferometer (MZI) configuration. In an MZI, light is first split into two arms and later recombined at the output port. Depending on the optical path length difference between the two arms, the output power can be maximized (in phase) or nulled (out of phase). The measured transmission spectrum through the on-chip MZI is shown in Figure 3(b), where we find the expected interference fringes with a free-spectral range (FSR) of 5.0 nm for a built-in optical path length difference of 100 μm. From the FSR we can calculate the group index of the optical mode to be around $n_g$=3.50, which agrees well with simulations. The MZI device provides a high extinction ratio of 25 dB, suggesting that the losses on the two arms are well balanced.

Another class of common photonic components in lightwave circuits involves resonant on-chip optical cavities. Here we demonstrate ring resonators. We fabricate devices with different coupling gaps in order to achieve critical coupling. First, we analyze ring resonators with a radius of 100 μm and a waveguide width of 900 nm. An optical micrograph of a fabricated device is shown in Figure 3(c).

The maximum extinction ratio in the transmitted optical signal is achieved for critical coupling conditions, which are reached when the optical power coupled into the ring matches the power dissipated inside the ring during one round trip. Under critical coupling conditions, the measured loaded optical $Q$ corresponds to half the intrinsic optical quality factor. For a near-critical coupling condition with a coupling gap of 300 nm, we find an extinction ratio of ~ 10 dB, as shown in the transmission spectrum in Figure 3(c). From fitting the resonance dip with a Lorentzian curve, we extract loaded optical $Q$ factors of 7,000. Using the relation $\alpha = 10\log_{10} e \cdot 2\pi n_g / Q_{int} \lambda$ (where $Q_{int}$ is the intrinsic quality factor, $\lambda$ is the wavelength and $n_g$ is the group index), we determine a propagation loss $\alpha$ = 44 dB/cm. When measuring devices with a larger coupling gap of 600 nm, the devices are operated in the under-coupled regime, and therefore the extinction ratio is smaller.

One important signature of ferroelectric materials is the dependence of dielectric properties on the ferroelectric domain structure. As grown, the BTO has its four in-plane tetragonal variants oriented at 90 degrees from one another. We demonstrate ferroelectric poling of these domains by applying an electric field to the ferroelectric along one arm of a BTO MZI device and monitor the output intensity for a wavelength at the quadrature point of the optical transmission spectrum, as indicated by the red circle on the MZI transmission spectrum in Figure 4(a). The length of the modulating arm is $L_{MZI}$=400 μm, and the spacing between the two electrodes is $g$=10 μm. The output intensity is measured as the electric field slowly sweeps with a peak-to-peak amplitude of 20 V at a frequency of $f$=10 Hz ($E_{off}$= $V_{off}$/$g$, where $g$ is the spacing between the two electrodes). As shown in Figure 4 (b), starting from zero applied field, the output intensity decreases because the applied electric field has an opposite sign to the polarization. At the coercive field for the ferroelectric in the waveguide, the polarization flips and the intensity of the output increases because the field direction is now aligned with the ferroelectric polarization. This hysteretic behavior, which has been reported in the literature in bulk devices[22-24], is observed as the sweep direction is reversed and reflects the ferroelectric behavior of the BTO localized in the waveguide region.

In the experiment to extract the effective Pockels coefficient of the BTO waveguide structures, we apply a sinusoidal electrical drive signal, as shown in the black curve in Figure 4(c), at a frequency of $f$ = 1 MHz with a DC bias $V_{bias}$= + 20 V to one arm of a MZI, which has an arm length of $L_{MZI}$ = 750 μm. The large DC bias $V_{bias}$ is used to produce a DC electric field greater than the coercive field of BTO so that all the ferroelectric domains are aligned and contribute most effectively to modulation. The sinusoidal electrical modulation signal is small in amplitude compared with $V_{bias}$. The modulated output light from the MZI is detected using a 10 MHz photoreceiver. Using the ratio of the amplitude of the output modulation (65 mV) to the optical transmission at the quadrature point (1.103 V), we calculate the voltage required to achieve a π phase shift $V_\pi$= 20 V, leading to a voltage length product of $V_\pi L$= 1.5 V·cm at a modulation frequency of 1 MHz. To ensure that we are indeed observing Pockels modulation instead of domain dynamics, we vary the DC bias voltage from –40 V to +40 V and perform the measurement each time at a quadrature transmission wavelength for each DC bias voltage. We find that the optical modulation amplitude saturates when $|V_{bias}| > 20V$. As a result, we believe with $V_{bias} = +20$ V, all the ferroelectric domains are fully poled, and thus we extract the effective Pockels coefficient of the BTO thin films to be $r_{eff} = 213 \pm 49$ pm/V by using $r_{eff} = \frac{\lambda g}{n_{eff}^3 \Gamma V_\pi L}$, where $n_{eff}$ is the effective mode index, $\Gamma$ is the electro-optic integral, $g$ is the spacing between the two electrodes (here $g=10$ µm), and $\lambda$ is the operation wavelength ($\lambda=1543.5$ nm). The error in the extraction of the effective $r_{eff}$ is computed by considering that the effective mode index $n_{eff}$ and EO overlap factor $\Gamma$ are affected by processing uncertainties in the silicon layer thickness ($\pm 15$ nm) and the BTO thickness ($\pm 10$ nm).

For many optical interconnect applications, it is desirable to operate the optical modulator at high speed to allow for high data transmission bandwidth. Figure 4(a) shows the experimental setup for measuring the bandwidth of the BTO modulators. In addition to the Mach-Zehnder interferometer shown as the device under test in the setup diagram, we study a resonant-type of optical modulator in a shape reminiscent of a "racetrack", as shown in Figure 1(c). This configuration has two straight sections, which provide a net modulation of the optical path length for TE-like modes propagating along the optimal [011] direction of the BTO.

Similar to the measurement at lower frequencies, the high frequency characterization is carried out with a bias field much greater than the coercive field ($|E_{bias}| = 2$ MV/m) applied on one of the MZI arms. Figure 4 (d) overlays the normalized electro-optic response amplitude $|S_{21}|$ for an MZI modulator with an arm length of 500 µm, and a racetrack modulator with a straight section length of 100 µm and loaded $Q$-factor of 6,000. Both spectra show flat frequency response below 100 MHz. This result confirms that the 1 MHz modulation shown in Figure 4(c) is due to the Pockels effect instead of slow domain dynamics, since transient domain dynamics, such as alignment and flipping, are known to have a much longer response time constant than 10 ns[25]. At high frequencies, the electro-optic response $|S_{21}|$ starts to roll off. An important metric of the $|S_{21}|$ spectrum is the frequency at which the optical modulation amplitude reduces to half of its value at low frequency, i.e. the 3 dB optical modulation bandwidth. From the data, the 3 dB modulation bandwidths are extracted to be $f_{3dB(MZI)}$ = 800 MHz and $f_{3dB(racetrack)}$ = 4.9 GHz for the MZI and racetrack modulator, respectively.

Because the length of the modulator (less than 1 mm) is small compared to the RF wavelength, both the electrodes for the MZI and racetrack modulators can be effectively modeled as a lumped-node circuit with a capacitor load. As shown in the optical micrographs of the MZI and microracetrack devices, the width of the gold electrodes next to the waveguide has been made as small as 1 μm to reduce the capacitance to the silicon layer to minimal. Using an LCR meter, we measure the electrode capacitance per unit length of electrode to be C = 10 pF/mm, resulting in an RC time constant limited 3 dB bandwidth of $f_{RC(MZI)}$ = 1.3 GHz for the MZI devices and $f_{RC(racetrack)}$ = 6.4 GHz for the racetrack devices. Both estimations are slightly greater than our measured bandwidths, suggesting factors such as RF loss in the metal also contribute to the roll-off. As shown in Figure 4 (d), the $|S_{21}|$ roll-off slope fits well to a -20 dB per decade trend predicted by a single-pole RC circuit model. We can see that the racetrack modulator has a greater bandwidth compared to the MZI modulator. This increased bandwidth is because the racetrack modulator has a shorter device length and hence smaller capacitance.

An experimental tool for evaluating the performance of a lightwave component is an eye-line diagram, where the signal from the receiver is repetitively sampled and applied to the vertical input of an oscilloscope, using the clock as the trigger. The resulting diagram, reminiscent of an "eye", can be used to evaluate the bandwidth and the efficiency of an optical modulator. Here, we apply a pseudorandom none-return-to-zero $2^{31}$-1 bit binary sequence onto the electrodes of the MZI modulator. The output light from the device is first amplified by an erbium-doped fiber-amplifier and then filtered and sent to a high-speed photoreceiver. For a modulating peak-to-peak voltage $V_{pp}$ = 6.6 V, we observe a clear eye opening at a data rate of 300 Mb/s with an extinction ratio of 3 dB, as shown in the inset of Figure 4(d).

The modulation efficiency will benefit if the optical loss in the BTO thin film can be further reduced to allow for both longer optical interferometers and microring resonators with higher *Q*-factor to be fabricated. Comparing the optical-Q factor ($Q$=7,000) demonstrated in our BTO microring resonator with the *Q*-factors of the silicon microring resonator in its early research history (Q~1,000)[26], we believe improvement gains can be realized by reducing the optical loss in the BTO thin films, which is currently limited by material and processing imperfections. Besides scattering from waveguide sidewall roughness, a major source of optical loss comes from the residual oxygen vacancies in the BTO thin film, which absorb light in the infrared[27]. These oxygen vacancies can arise both during the MBE process and during the a-Si deposition, which is performed in a reducing atmosphere for hydrogenation. *In-situ* oxygen plasma annealing inside the growth chamber and postgrowth annealing in a highly reactive oxidizing environment, such as ozone, can aid in reducing the residual oxygen vacancy density. In addition, a hydrogen barrier layer[28], such as a thin layer of alumina and silicon nitride encapsulating the ferroelectric thin film, can be useful to prevent oxygen vacancies from forming inside the oxide thin films during the subsequent a-Si deposition.

In conclusion, we have shown the direct integration of ferroelectric BTO on silicon waveguide structures exhibiting a strong linear electro-optic effect with an effective Pockels coefficient of 213 ± 49 pm/V. The ferroelectric poling of epitaxial BTO is confirmed through on-chip electro-optic measurements. We use a horizontal slot waveguide structure to demonstrate the commonly used passive photonic components, including on-chip interferometers and microring resonators. The horizontal slot waveguide structure provides a large overlap between electric and optical fields, which is important for building active BTO electro-optic devices. The modulation speed can be much improved using a well-studied traveling-wave design[29, 30]. Beyond optical communication applications, our research may open new routes towards integrating noncentrosymmetric crystalline oxides for realizing on chip many exciting nonlinear and non-classical optical functionalities such as wavelength conversion[31] and nonclassical light source[32].

## Acknowledgement

The authors acknowledge funding support from the NSF through MRSEC DMR-1119826 (CRISP). F.J.W. and C. H. A. also acknowledge support from NSF through DMR-1309868. The authors thank Dr. Michael Rooks and Michael Power for assistance in device fabrication.

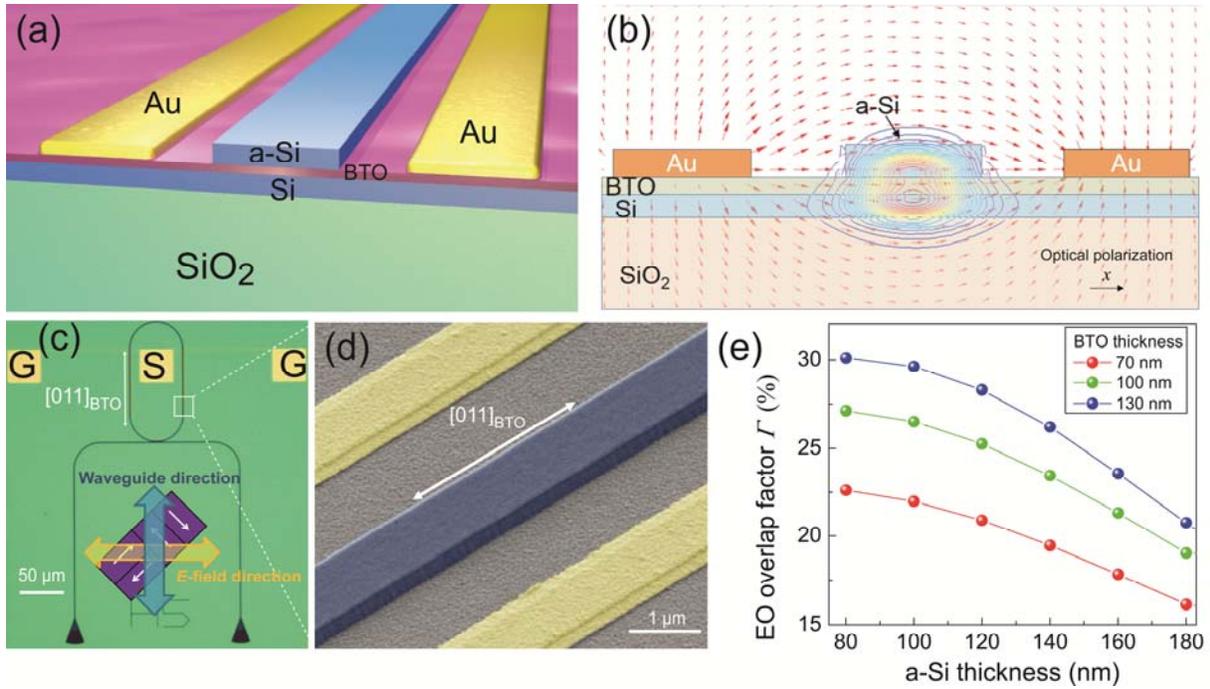

**Figure 1. Integrated barium titanate (BTO) electro-optic modulators**. (a) A three-dimensional schematic of the BTO modulator fabricated on silicon-on-insulator substrates. (b) Fundamental optical TE-like mode profile (plotted in contour lines of $E_x$, the in-plane optical polarization) and electric field distribution (in arrows). (c) An optical micrograph of a BTO micro-racetrak optical modulator. The electrodes (ground-signal-ground, GSG) are patterned in close proximity to the modulating waveguides. The optical waveguide is parallel to BTO's [011] lattice direction. The inset schematic illustrates the four in-plane domain variants with their polarziation directions represented in white arrows and their relations to the waveguide (shown as the yellow arrow) and E-field orientation (shown as the blue arrow). (d) A false-color scanning electron micrograph of the amorphous silicon waveguide (in blue) and the gold electrodes (in yellow.) (e) Numerically calculated electro-optic overlap factor ($\Gamma$) as a function of a-Si thickness and BTO thickness.

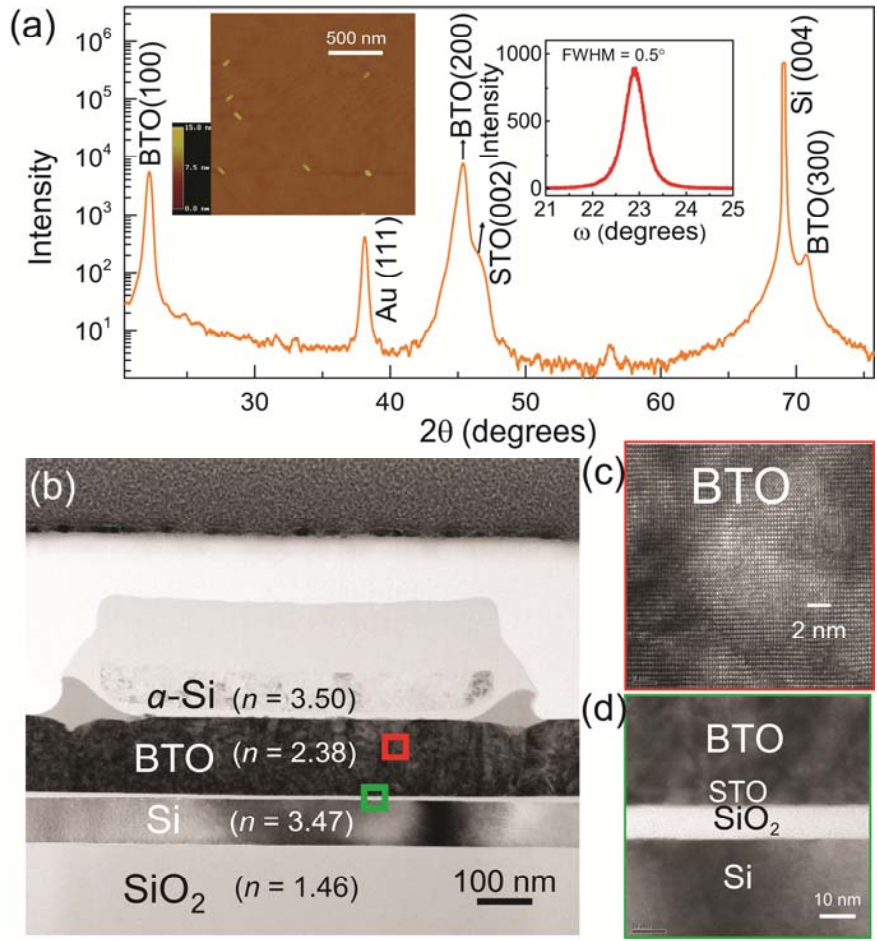

**Figure 2. Materials analysis of epitaxial BTO on SOI**. (a) θ-2θ X-ray diffraction pattern from a fully processed Si-waveguide integrated BTO modulator, which shows Bragg reflections from BTO(100), BTO(200), BTO (300), STO (002), Si(004) and Au (111). The left inset is a 5 μm ×5 μm atomic force micrograph on the surface of the as-grown 80 nm BTO on SOI. The right inset shows a rocking curve about the BTO(200) Bragg reflection. (b) High-resolution transmission electron micrograph of a cross sectional area of the a-Si/BTO/SOI horizontal slot waveguide. The micrographs at higher magnification for (c) show lattice fringes, indicating the single crystal nature of the BTO thin film, while (d) reveals the strontium titanate (STO) buffer. The amorphous $SiO_2$ layer shown in (d) is the result of oxidation of the silicon after oxygen annealing.

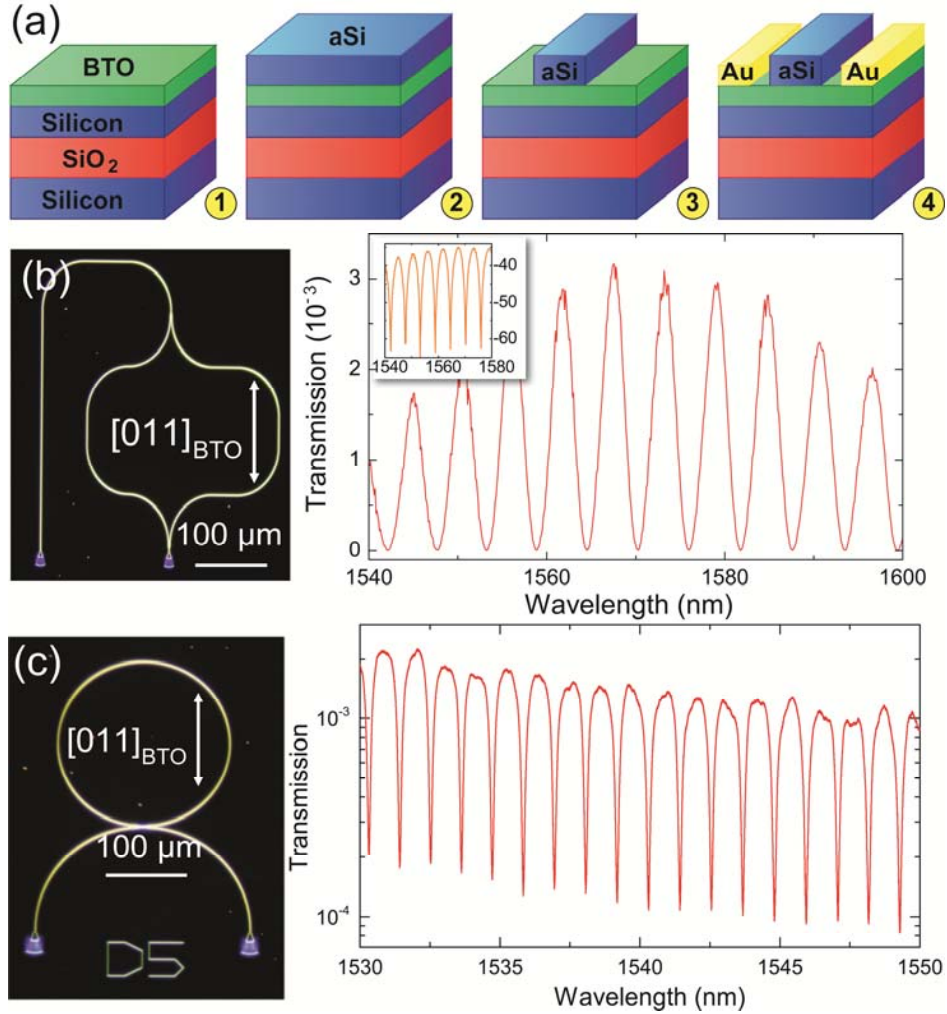

**Figure 3. Si-integrated BTO waveguide devices.** (a) Fabrication flow for the integrated BTO modulators. (a1) Epitaxial growth of an 80 nm-thick film of BTO on STO/SOI. (a2) Plasma-enhanced chemical vapor deposition of 110 nm-thick amorphous silicon on BTO. (a3) Nanolithography and patterning of the a-Si waveguides. (a4) Lithography and patterning of the Ti/Au electrodes. (b) Optical micrograph of a BTO on SOI Mach-Zehnder interferometer (MZI). The straight segment of the MZI arm is aligned to BTO lattice's [011] direction. The optical transmission is shown in both linear (right) and log scale (inset). The extinction ratio of the MZI is 25 dB. (c) Optical micrograph of a microring resonator slot waveguide. BTO lattice's [011] direction is shown as reference. The radius of the ring shown is 100 μm, and the typical loaded optical Q-factor is 7,000. The optical transmission is shown on the right.

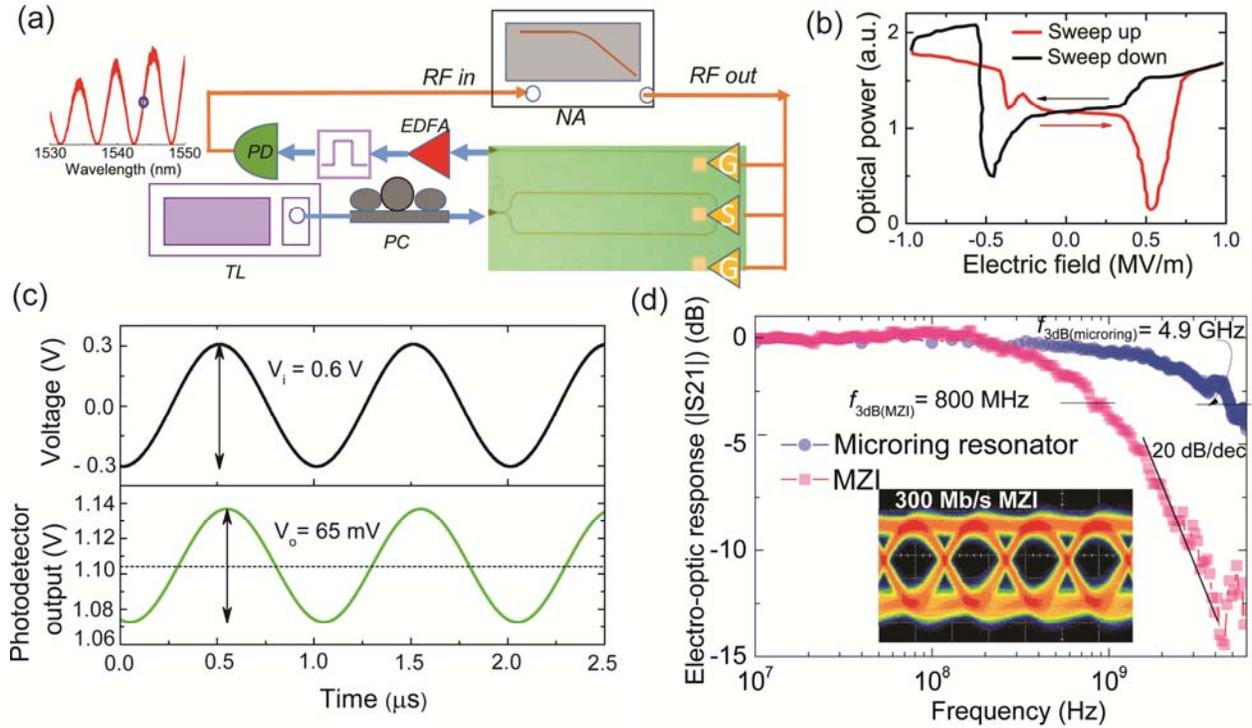

**Figure 4. Electro-optic characterization of the Si-waveguide integrated BTO MZI modulators.** (a) Experimental setup for characterizing the electro-optic properties of a BTO-MZI type modulator. TL: tunable laser. PC: polarization controller. EDFA: Erbium-doped fiber amplifier. PD: High-speed photodiode. NA: network analyzer. The wavelength of the input light is tuned at the quadrature point of the MZI transmission, as shown on the left (blue circle on red MZI optical transmission curve). (b) DC optical transmission as a function of the lateral eletric field applied across the electrodes. For this measurement, the electric field is swept with a sawtooth pattern at a frequency of 10 Hz. A hysteresis is observed as the sweep direction is reversed. (c) Photodetector output waveform (bottom curve in green) when a 1 MHz sinusoidal RF signal with peak-to-peak voltage of $V_i$=0.6 V (upper curve in black) is applied across the electrodes. From the relative amplitude of the modulation signal $V_o$ and the quadrature transmission, the switching voltage $V_\pi$ can be inferred. (d) Normalized electro-optic response ($|S_{21}|$) in dB

as a function of the modulating frequency in logarithmic scale. The -3 dB electro-optic bandwidths for the Mach-Zehnder interferometer (MZI) and the racetrack resonator are 800 MHz and 4.9 GHz, respectively. The inset is a 300 Mb/s non-return-to-zero eyeline diagram measured on the integrated BTO MZI modulator.